# Cascaded Four-Wave Mixing on Quantum Paraelectrics for On-chip Cryogenic Microcombs

[#]Harikrishnan Sundaresan, [#]Prasad Muragesh, and Madhu Thalakulam[*]

Indian Institute of Science Education and Research Thiruvananthapuram, Kerala, India 695551

## Abstract

Here, we demonstrate an on-chip cryogenic microwave frequency-comb on a planar superconducting resonator fabricated on SrTiO3, a quantum paraelectric material. The material's Kerr-type microwave nonlinearity arising from the quantum paraelectric phase an underexplored state, enables pronounced resonance shifts, Duffing-like bifurcations and comb generation via cascaded four-wave mixing. Our results establish SrTiO3 resonators as a low power, cryogenically compatible platforms for nonlinear microwave photonics, with applications in scalable quantum control and on-chip frequency synthesis.

[#] Equal Contribution

[*] madhu@iisertvm.ac.in

*Introduction*—Miniaturisation and scalability have driven semiconductor technology since the invention of the transistor in 1959 [1]. The very same mottos are now equally relevant for advances in quantum technologies too. Currently, quantum circuits, such as semiconducting and superconducting quantum processors and quantum electrical metrology circuits, operating at sub-kelvin temperatures, heavily use microwave pulses for initialization, manipulation and readout of states [2,3]. These are accomplished with the help of bulky conventional microwave electronics posing a constraint when it comes to scalability, wiring complexity, thermalization and reduction in noise and decoherence rates [4,3]. For photonic quantum technologies, frequency combs provide a convenient and stable frequency source for a variety of applications [5] including spectroscopy [6], optical clocks [7], frequency metrology, and frequency synthesis [8–10]. Recent developments focus on the scalability aspects, exploiting integrated photonic devices capable of generating mode-locked frequency combs either using microresonators defined on a nonlinear medium [8,11,12] or using mode-locked lasers [7]. Akin to this, an on-chip implementations of microwave frequency combs satisfying low power requirements, phase-locking, and cryogenic operation could greatly sole many of the teething problems, though largely missing except for a few isolated efforts [13–16]. Microwave frequency combs have been realized on superconducting devices based on Kerr nonlinearities found in Josephson-junctions [13,17,18] (JJ) and high kinetic inductance (KI) materials [15,19] such as NbN and NbTiN, as well as magnonic [20,21], phononic [22–24] and electromechanical [19,25] systems. Among these, JJ- and KI-based frequency combs appear to be the most robust demonstrations at cryogenic temperatures, with wide frequency spans, low power operation and phase coherence. But they are not without drawbacks, as JJs are sensitive to magnetic fields and the FSR of the combs on these platforms are dependent on multi-mode coupling, and thus tuneability is limited. Additionally, the feasibility of JJ-based devices hinges on their intense fabrication requirements. The comb span of electromechanical [25–27], phononic [22,24] and magnonic [21] combs are limited to a few spectral lines and are generally not operated at cryogenic temperatures with some exceptions [19,27]. These limitations motivate the exploration of alternative nonlinear platforms for on-chip microwave frequency combs compatible with cryogenic low-power operation.

Systems with intrinsic second or third order nonlinearity in the microwave regime at cryogenic temperatures are a prerequisite for the generation of microwave frequency combs. In this regard, quantum paraelectric materials such as strontium titanate [28] ($SrTiO_3$ or STO) and potassium tantalate ($KTaO_3$ or KTO) due to their strong, intrinsic, and tuneable dielectric nonlinearity [29–31] at cryogenic temperatures are potential candidates. Strontium titanate ($SrTiO_3$ or STO) has a centrosymmetric perovskite crystal structure exhibiting a pronounced third-order nonlinear dielectric response at low temperatures, arising from soft transverse phonon modes [30–32]. Unlike conventional electro-optic materials such as $LiNbO_3$ and Si, this nonlinearity persists into the microwave regime [30,33,34]. This third order Kerr nonlinearity lends itself to many potential applications for compact cryogenic quantum circuits which have recently found growing interest [34]. STO-based varactors have been used for frequency tuning and impedance matching of readout circuits [35,36], and recent theoretical proposals to use quantum paraelectric materials for three-wave mixing based parametric amplifiers [33], as well as efforts to reduce intrinsic material losses using thin-film STO [37] indicate ongoing efforts to implement practical nonlinear microwave devices using STO. In addition, STO-based devices are, in principle, compatible with magnetic fields offering additional flexibility for hybrid circuit integration. The Kerr nonlinearity found in STO enables four-wave mixing dynamics desirable for the generation of microwave frequency combs [15,19], which are usually demonstrated at optical frequencies using microresonators [11,38,39].

In this work, we report the generation of coherent microwave frequency combs via cascaded four-wave mixing in a superconducting CPW resonator defined on STO. The 4WM processes, further enhanced by the cavity, are enabled by the presence of Kerr nonlinearity in the resonator, which has until now not been experimentally reported in STO based devices. We characterize the Kerr nonlinearity in the system by measuring the power-dependent resonance frequency shift and observing the onset of jumps characteristic of Duffing-like bifurcations at high drive powers. The estimated self-Kerr coefficient of $\sim$1 mHz per photon is comparable to those reported for KI-based nonlinear resonators. Building on this nonlinear response, we apply two coherent pump tones centred around resonance with drive powers much higher than the bifurcation threshold. The resulting frequency comb consists of more than a thousand phase-locked comb lines with line-to-line flatness of ~0.1 dB. Furthermore, the FSR of the comb is tuneable by adjusting the spacing between the two supplied pump tones. These results establish superconducting STO-based resonators as a viable and distinct new class of nonlinear superconducting devices, capable of producing rich nonlinear phenomena such as phase-locked frequency combs. Our approach relies on an intrinsic material nonlinearity that is stable, low-loss, and compatible with cryogenic integration and electric field tuning. The demonstration of coherent microwave frequency combs suggests immediate applications in on-chip frequency synthesis and multiplexing of quantum control signals. More broadly, this work opens the door to leveraging quantum paraelectricity for scalable, versatile nonlinear microwave photonics in superconducting circuits.

*Kerr nonlinearity of STO in the quantum paraelectric state*—By embedding a superconducting aluminium coplanar waveguide (CPW) resonator on STO, the nonlinear dielectric response can be enhanced and is accessible through microwave reflection or transmission measurements. Figure 1 introduces the device architecture and establishes the nonlinear response underlying frequency-comb generation. As illustrated in Fig. 1(a), the superconducting resonator, made of Aluminium, with a resonance frequency of 474 MHz in the low-power linear regime is fabricated on an STO substrate and operated at a temperature of $\sim 100$ mK. In this temperature range, STO is in the quantum paraelectric state [29,30] with a field-dependent permittivity, resulting in an intrinsic third order $\chi^{(3)}$ nonlinearity. While broken inversion symmetry in STO can induce a second-order $\chi^{(2)}$ nonlinearity, our experimental focus in this work is on the third-order $\chi^{(3)}$ Kerr-like nonlinearity. By employing an intra-cavity dual-tone pumping scheme centred around the fundamental resonance, we investigate four-wave mixing processes observed in the cavity. The system as shown in Fig. 1 A can be modelled as a nonlinear resonator, driven by two coherent microwave pump tones ($f_0$ and $f_1$) fixed around the fundamental resonance with the detuning $f_1 - f_0$ determining the free spectral range (FSR) of the comb. The strong Kerr-like interaction within the STO substrate facilitates both degenerate and non-degenerate four-wave mixing (4WM) processes, as shown in the illustrations in Fig. 1 A. In this regime, the pump tones mix with one another and with the resulting sidebands, seeding a cascaded mixing process where energy is redistributed across a broad spectral range going far beyond the cavity bandwidth, eventually manifesting as a phase-coherent frequency comb. The Hamiltonian of the resonator can be written as

$$H = \hbar\omega_0(a^\dagger a) + \hbar\frac{K}{2}(a^\dagger a)^2$$

where $\omega_0$ and $K$ are the resonance frequency and self-Kerr coefficient of the fundamental mode, with $a^\dagger$ and $a$ being the creation and annihilation operators [19,33]. The first term is the Hamiltonian corresponding to the simple harmonic oscillator and the second term introduces drive dependent nonlinear anharmonicity. We characterize the nonlinearity in our resonator by experimentally measuring the self-Kerr coefficient. The nonlinear cavity response is characterized through reflection measurements, as shown in Fig. 1(b). Details of the experimental setup is provided in Supplementary Information SI-1. As the input power is increased, the resonance frequency shifts towards lower frequencies, followed by a jump phenomenon indicative of metastable states above $-46$ dBm input power. This behaviour is a hallmark of a Duffing-type nonlinear oscillator and reflects the strong Kerr-like nonlinearity induced by the STO substrate [19,40,41]. Fig. 1(c) shows representative line cuts (highlighted in Fig. 1(b) of the reflection coefficient at selected powers, highlighting the power-dependent resonance shift and metastability around the resonance. The extracted shift in the resonance scales linearly with the average intracavity photon number $\tilde{n}$ (inset of Fig. 1(c)), consistent with a Kerr-type nonlinearity (see Supplementary Information SI-2 for details of Kerr nonlinearity calculation). The self-Kerr coefficient given by the slope $\Delta\omega/\Delta\tilde{n}$ of the linear fit for our device is $(-10.20 \pm 0.35) \times 10^{-4}$ Hz per photon. These measurements establish the presence of a sufficient Kerr nonlinearity on STO in the microwave regime, providing the physical foundation for cascaded four-wave mixing and broadband frequency comb generation discussed later. While theoretical proposals have suggested the potential of quantum paraelectrics for parametric three-wave mixing based amplification [33], a quantitative experimental determination of the Kerr coefficient in an STO-based superconducting resonator has remained absent from the literature until now.

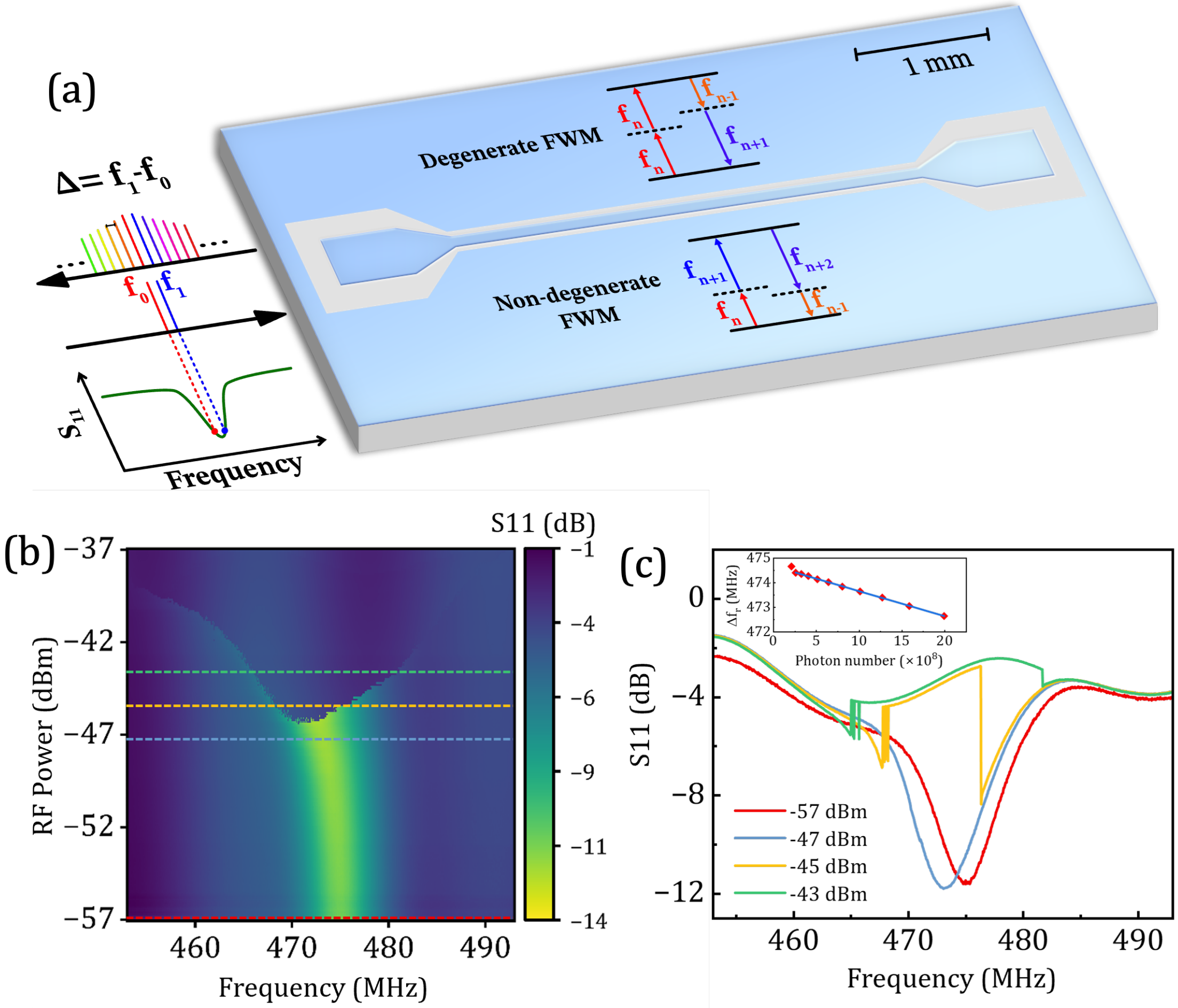

**Figure 1.** **(a)** Schematic representation of the device architecture featuring a superconducting Aluminum coplanar waveguide (CPW) resonator fabricated on a Strontium Titanate (STO) substrate. The illustrations depict the intra-cavity dual-tone pumping scheme and the resulting degenerate and non-degenerate four-wave mixing (4WM) processes. **(b)** Heatmap of the reflection coefficient ($S_{11}$) as a function of frequency and RF input power, illustrating the nonlinear evolution of the resonance. **(c)** Representative line cuts from (b) at selected input powers (from -57 dBm to -43 dBm) showing a downward frequency shift and the onset of Duffing-like bifurcation and metastability above -46 dBm. **Inset:** Measured resonance shift ($\Delta f_r$) versus the average intracavity photon number $\tilde{n}$. The linear fit confirms a Kerr-type nonlinearity with a self-Kerr coefficient of $(-10.20 \pm 0.35) \times 10^{-4}$ Hz per photon.

*Coherent electrically tuneable microwave frequency comb generation*—Demonstrating the presence of Kerr nonlinearity in our system, we now describe the emergence of coherent microwave frequency-comb from cascaded four-wave mixing. To trigger the comb generation, we apply two continuous-wave pump tones, $f_0 = 474\ MHz$ and $f_1 = 474.1\ MHz$, around the cavity resonance (highlighted in left inset of Fig. 2(a)). The frequencies corresponding to the comb lines are indexed by $f_n = f_0 + n\Delta$ where $n$ is an integer, and $\Delta = f_1 - f_0$ is the free spectral range (FSR). At drive powers $P_{f0} = P_{f1} \approx -36.5$ dBm significantly exceeding the bifurcation threshold ($\sim -46$ dBm), the resonator enters a regime of efficient, cascaded four-wave mixing (4WM). As shown in Fig. 2(a), this results in a dense frequency comb with FSR = 100 kHz consisting of over a thousand spectral lines. We observe a comb spanning more than 150 MHz—extending from roughly 400 MHz to 550 MHz. This is much higher than the cavity bandwidth of $\sim$10 MHz. Supplementary Information SI-3 shows the generation of a similar frequency comb at higher frequencies using a resonator whose fundamental resonance is around 983 MHz.

The spectral envelope is remarkably well-defined; the right inset to Fig. 2(a) shows a magnified view of the comb lines in the highlighted region (enclosed in green rectangle), demonstrating high uniformity, with line-to-line variations of approximately 0.1 dB. This flatness is critical for applications in multiplexed quantum control, as it ensures uniform signal strength across multiple channels. Furthermore, high-resolution measurements confirm that individual comb lines have a spectral width of less than 1 Hz, limited only by the resolution bandwidth of the measurement setup [15,17].

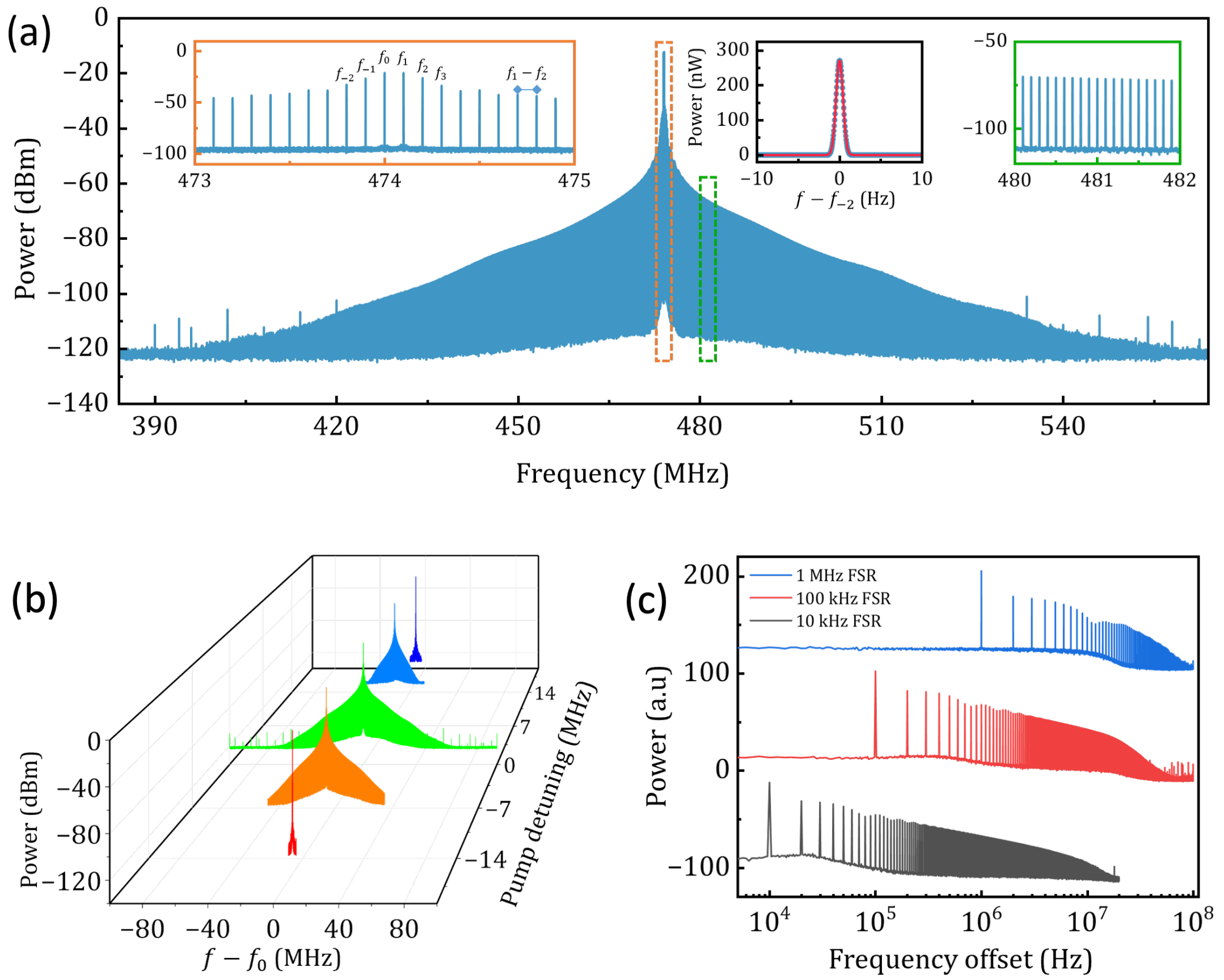

**Figure 2.** **(a)** Broadband microwave frequency comb spanning over 150 MHz. **Left inset:** Magnified view of the central region (highlighted orange rectangle) showing the two primary pump tones ($f_0$ and $f_1$) and the indexed comb lines. **Right inset:** High-resolution view (highlighted green rectangle) demonstrating a flat spectral envelope with approximately 0.1 dB line-to-line uniformity. **Central inset:** Spectral resolution measurement of an individual comb line, limited to $< 1\ Hz$ by the instrument resolution bandwidth. **(b)** 3D plot illustrating the cavity-enhanced nature of the process. Comb generation is maximized when the pump pair is centred on the resonance and diminishes as the pumps are detuned outside the resonator bandwidth. **(c)** Demonstration of FSR tunability over three orders of magnitude (10 kHz, 100 kHz, and 1 MHz) achieved by adjusting the pump detuning.

The role of the resonator in mediating these 4WM interactions is explicitly shown in Fig. 2(b). By sweeping the pump pair across the resonance frequency while maintaining a fixed FSR of 100 kHz, we observe that comb generation is sensitive to the pump-cavity detuning. The comb span and power are maximized when the pumps are centred on the resonance and diminish as the drive tones move outside the cavity bandwidth. However, the comb continues to persist with a reduced span even when the drive tones are slightly detuned from the cavity resonance. This is because the drive tones give rise to 4WM products that fall within the resonator's bandwidth. These sidebands— enhanced by the cavity— function as secondary pumps which trigger the cascaded 4WM process manifesting as a frequency comb with a reduced span. The superconducting resonator acts as a high-confinement volume that resonantly enhances the intracavity microwave fields. By recirculating the pump photons, the cavity effectively lowers the external power threshold required to trigger cascaded 4WM. This enables the synthesis of a broadband coherent frequency comb spanning far beyond the cavity bandwidth and using relatively low input powers that would otherwise yield only a few mixing products in a non-resonant geometry.

A defining feature of this STO-based device is the control over the comb's FSR. In contrast to combs generated in multimode resonators where the FSR is constrained by the physical geometry and multi-mode coupling of the device [14,22], the FSR here is strictly governed by the spacing of the primary pump tones $f_0$ and $f_1$. We demonstrate this flexibility in Fig. 2(c), by varying the pump detuning to produce combs with FSRs spanning three orders of magnitude: 10 kHz(black), 100 kHz (red), and 1 MHz(blue). Thus, the FSR of the comb is tuneable from anywhere between a few Hz to a few MHz, limited by the cavity bandwidth as discussed earlier. Notably, the total number of spectral lines remains largely conserved across these cases, which implies that the power distribution between the comb lines depends on the nonlinear coupling strength unless the pumps are far outside the cavity bandwidth, where cavity-enhancement is absent. This tuneability allows for "on-the-fly" reconfiguration of the comb spacing, a capability that is elusive in JJ-based devices where FSR is often fixed by multi-mode coupling [14].

*Power dependence*—We now investigate the power dynamics of the pump tones with respect to each other in determining the threshold of comb formation. Figure 3 illustrates the dependence of the cascaded 4WM process on both the absolute pump power and the power balance between the two pumps. The transition from the generation of a few mixing products to a broad, dense, comb state is explored in Fig. 3(a), by simultaneously sweeping the power of both pumps $P_{f0}$ and $P_{f1}$. At low input powers (-50.5 dBm), the spectrum remains sparse, with only the two primary pump tones and weak first-order sidebands visible. As the power is increased far beyond the bifurcation threshold to around -36.5 dBm, we observe a sharp transition to a broad comb. The number of sidebands abruptly increases, resulting in a dense comb that spans over 150 MHz. This abrupt onset suggests that the comb state is a collective nonlinear phenomenon triggered once the intracavity field density is sufficient to sustain high-order cascaded mixing. Interestingly, as the power is further increased beyond the optimal comb formation point (≈ 30.5 dBm), the comb span begins to narrow and line to line variation increases. This likely indicates an increased phase mismatch induced by the significant power-dependent resonance shift.

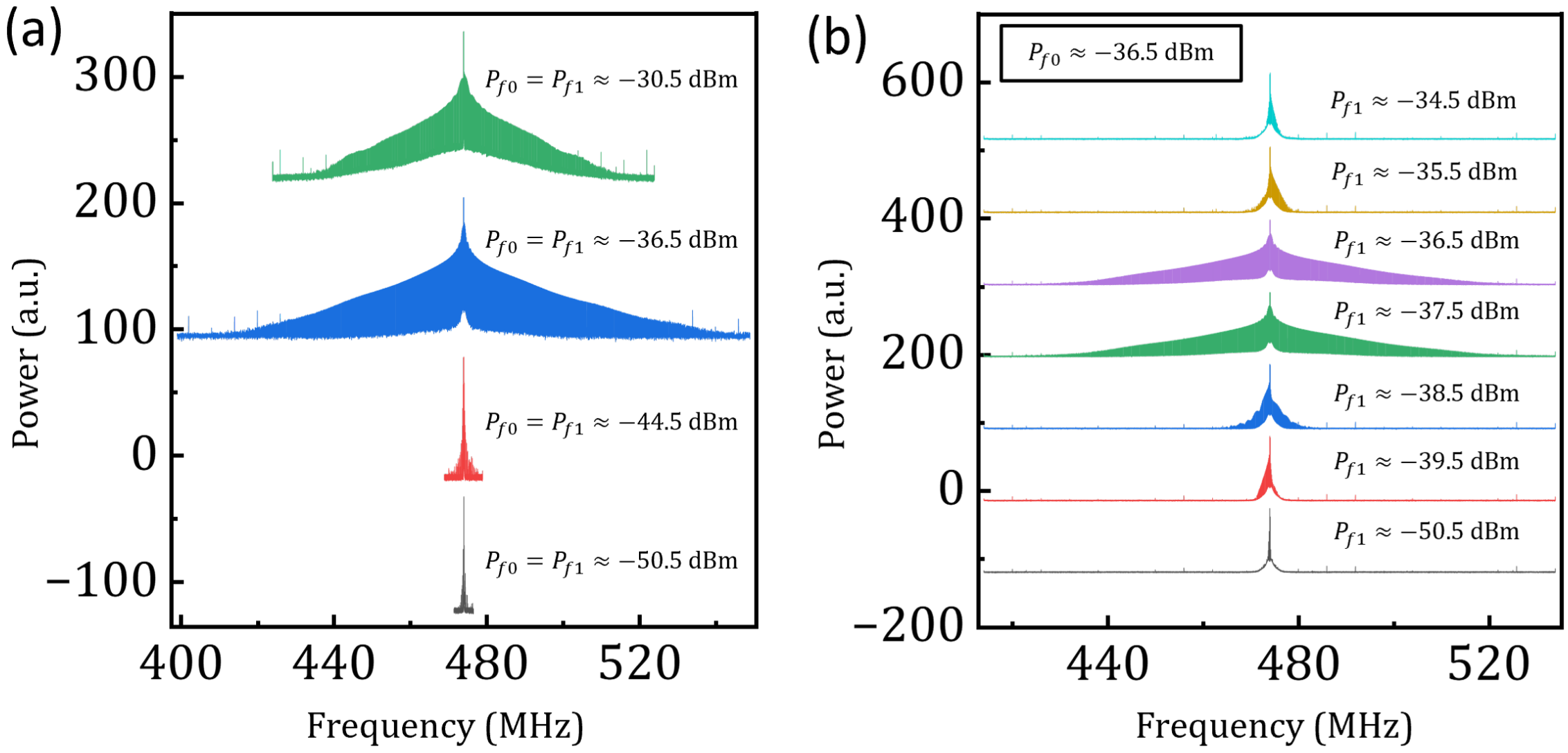

**Figure 3.** **(a)** Evolution of the frequency comb as the power of both pumps ($P_{f0}$ and $P_{f1}$) is swept simultaneously. A sharp transition from sparse mixing products on either side of the pumps to a dense broadband comb is observed as power exceeds the bifurcation threshold, optimal near -36.5 dBm. Further power increases result in a narrowing comb span due to increased phase mismatch. **(b)** Frequency comb dependence on pump power balance. With $P_{f0}$ fixed at -36.5 dBm, a broadband comb only emerges when $P_{f1}$ is nearly balanced with respect to $P_{f0}$. Power imbalance disrupts the cascaded energy redistribution, leading to a collapse of the spectrum into a sidebands on either of the pumps.

We also find that the stability of these combs is dependent on the power balance between the two pumps, as shown in Fig. 3(b). When $P_{f0}$ is held fixed at $-36.5$ dBm (near the optimal point) and $P_{f1}$ is varied, the broadband comb structure only emerges when the tones are nearly equal in amplitude ($P_{f0} = P_{f1}$). Deviation from this balance leads to a rapid reduction from the broadband symmetric spectrum into a few mixing products on either side of the pumps. This imbalance in pump power disrupts the cascaded energy redistribution from the pumps to the sidebands, preventing the generation of higher-order sidebands.

*Phase coherence and time-domain analysis*—We confirm the phase coherence of the generated frequency comb by analysing its time-domain signal. Fig. 4(a) shows the temporal signal recorded on a 5 GS/s digital storage oscilloscope when two coherent pump tones (each at -36.5 dBm) are applied symmetrically around the resonance. The signal consists of a periodic sequence of pulses, with a repetition rate equal to the FSR set by the pump detuning. The non-sinusoidal and asymmetric pulse-shape comes from the summation of over a thousand sinusoids corresponding to the spectral lines, each associated with a well-defined phase [42]. The Fourier transform of the time domain signal, shown in Fig. 4(b), reproduces the frequency comb structure observed via spectrum analyser in Fig. 2(a), including the same bandwidth and line spacing. A magnified view, shown in inset to Fig. 4(b), confirms that the comb consists of nearly equal-amplitude, evenly spaced lines, indicating a flat spectral envelope near the centre of the comb.

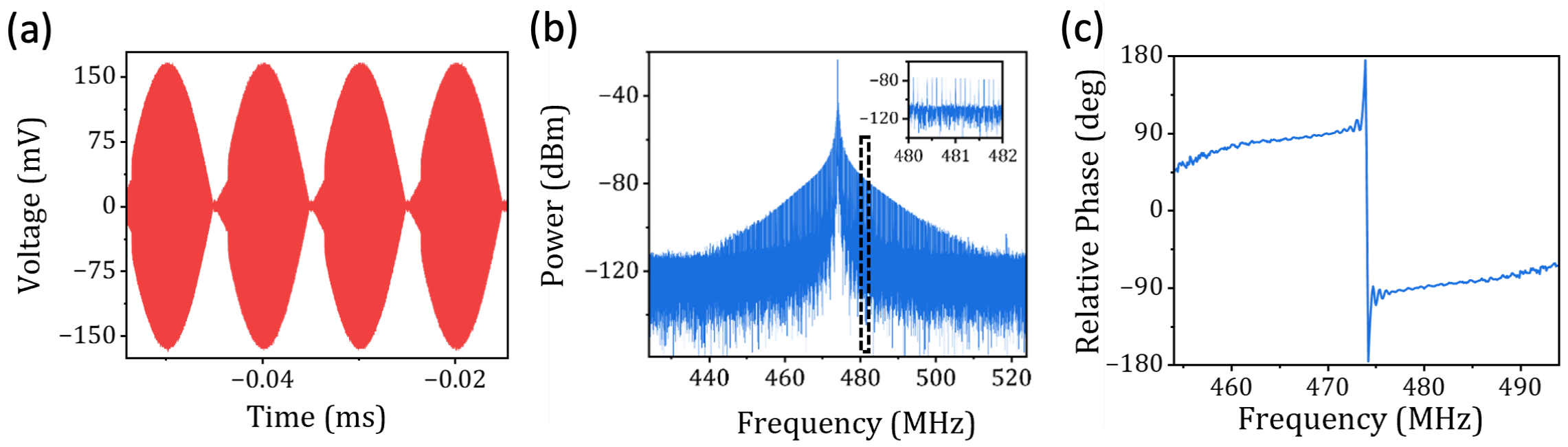

**Figure 4. (a)** Time-domain voltage signal of the generated comb showing a periodic pulse train with a repetition rate dictated by the FSR (here 100 kHz). The non-sinusoidal pulse shape arises from the coherent summation of over a thousand phase-locked spectral lines. **(b)** Fourier transform of the temporal signal, reproducing the spectral structure and 100 kHz line spacing. **Inset:** Magnified view of the highlighted black region confirming the flat spectral envelope near the comb centre. **(c)** Extracted relative phases of the comb lines across the measured bandwidth. The fixed, stable phase relationship confirms that the spectral components are phase-locked.

To quantify the phase coherence across the spectrum, we extract the relative phases of the comb lines from the Fourier-transform of the time-trace. The extracted phases in Fig. 4(c) exhibit a fixed relationship between the comb lines across the measured bandwidth confirming that the spectral components are phase-locked. Furthermore, this phase relationship remains stable over extended periods (see Supplementary Information SI-4 for details), thus demonstrating the long-term stability of the comb.

*Conclusions*—In summary, we have demonstrated the generation of phase-coherent microwave frequency combs using the intrinsic third-order nonlinearity of a quantum paraelectric STO-based superconducting resonator. By leveraging the voltage-tuneable Kerr response of the STO substrate at millikelvin temperatures, we achieved a cascaded four-wave mixing process that yields over a thousand stable comb lines, compared to the few lines reported in previous magnonic or electromechanical platforms which are based on similar nonlinearities. Unlike JJ and KI-based combs, where the FSR is often constrained by fixed inter-mode coupling, our platform offers tuneable FSR through pump-tone detuning. Additionally, frequency comb generation in our device spans several linewidths beyond the cavity resonance. The line-to-line flatness and long-term phase stability highlight the suitability of STO-based nonlinear resonators for precision microwave applications such as frequency synthesis, multi-tone readout, and coherent signal generation. Such a system could potentially replace dozens of individual microwave control lines in a dilution refrigerator, addressing the cabling bottleneck that currently limits large-scale qubit integration. The device is also compact owing to the very high dielectric constant of STO and relatively easy to fabricate compared to JJ-based devices. Our results establish STO-based superconducting devices as a robust alternative platform for nonlinear microwave photonics compared to JJ- and KI-based devices, also suggesting potential applications for parametric amplifiers [33], bifurcation amplifiers, non-reciprocal devices, mixers/modulators, etc.

**Acknowledgements**: MT acknowledge funding support from the National Quantum Mission, an initiative of the Department of Science and Technology, Government of India, and PM acknowledges UGC for fellowship.

**Author contributions**: MT conceived the problem, PM fabricated the devices, HS and PM performed the measurements, HS analysed the data, and HS & MT co-wrote the manuscript.

**Competing interests:** The authors declare no competing interests.

**Data and code availability:** All data and code used in the analysis and supporting data are available from the authors upon request.

# Supplementary Information

## Cascaded Four-Wave Mixing on Quantum Paraelectrics for On-chip Cryogenic Microcombs

[#]Harikrishnan Sundaresan, [#]Prasad Muragesh, and Madhu Thalakulam[*]

Indian Institute of Science Education and Research Thiruvananthapuram, Kerala, India 695551

# SI-1. Measurement Setup

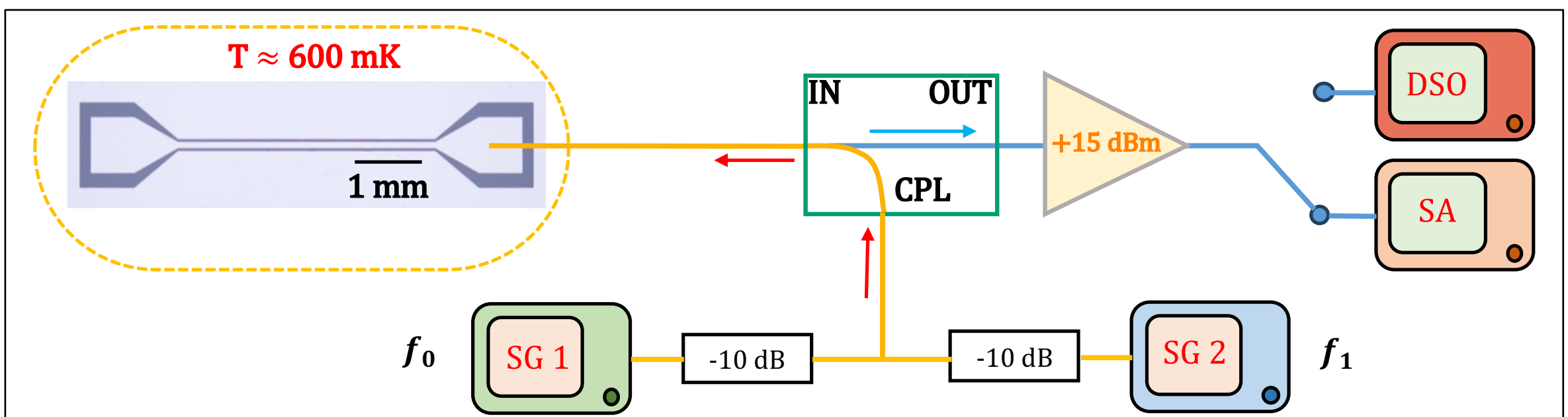

Figure S1: Measurement setup for the experiment. Optical micrograph of a similar device used in the experiment is highlighted by the dotted yellow oval. Two microwave signal generators (SG 1 and SG 2) are connected through a tee with -10 dB attenuators to prevent crosstalk, via the directional coupler to the device placed in the still stage of the dilution refrigerator. The reflected signal from the device is amplified at room temperature with a low noise amplifier of gain +15 dB and connected to spectrum analyser (SA) for frequency domain measurements and to a digital storage oscilloscope (DSO) with 5GS/s sample rate and 1 GHz bandwidth for time domain measurements.

# SI-2. Calculating Kerr Nonlinearity

The Kerr nonlinearity of the superconducting CPW resonator on STO can be characterized by the measuring the power dependence of its resonance. The Hamiltonian of the resonator is given by

$$H = \hbar\omega_0(a^\dagger a) + \hbar\frac{K}{2}(a^\dagger a)^2 \qquad (1)$$

[#] Equal Contribution

[*] madhu@iisertvm.ac.in

where $\omega_0$ and $K$ are the low input power resonance frequency and self-Kerr coefficient of the fundamental mode, with $a^\dagger$ and $a$ being the creation and annihilation operators. The Kerr coefficient $K$ is given by $\Delta\omega/\Delta n$ where $\Delta\omega$ is the shift in resonance for a change in the intracavity photon number $\Delta n$. The intracavity photon number can be estimated if we know the internal and coupling Q factors of the resonator, which are extracted from the complex S11 modelled as

$$S_{11} = 1 - \frac{2Q_l/Q_c}{1 + 2iQ_l(\omega - \omega_r)/\omega_r} \tag{2}$$

Where $Q_l$, $Q_c$ are the loaded and coupling Q-factors and $\omega_r$ is the angular resonance frequency. We fit Eqn. (2) using the circle-fit method [1,2] to find $Q_l$, $Q_c$, $\omega_r$ and the internal Q-factor $Q_i$. Fig. S2(a) and (b) shows the phase and magnitude of $S_{11}$ respectively, around the resonance at $P_{chip} = -57$ dBm along with the fits to Eqn. (2) (dotted red line). We note that the fit to the resonance is not exact as the resonator is over-coupled ($Q_c < Q_i$), leading to an increase in the systematic uncertainty in the fitting [2] and an increase in the uncertainty of $Q_i$. This uncertainty has been taken into account in the following calculations of the intracavity photon number and Kerr coefficients.

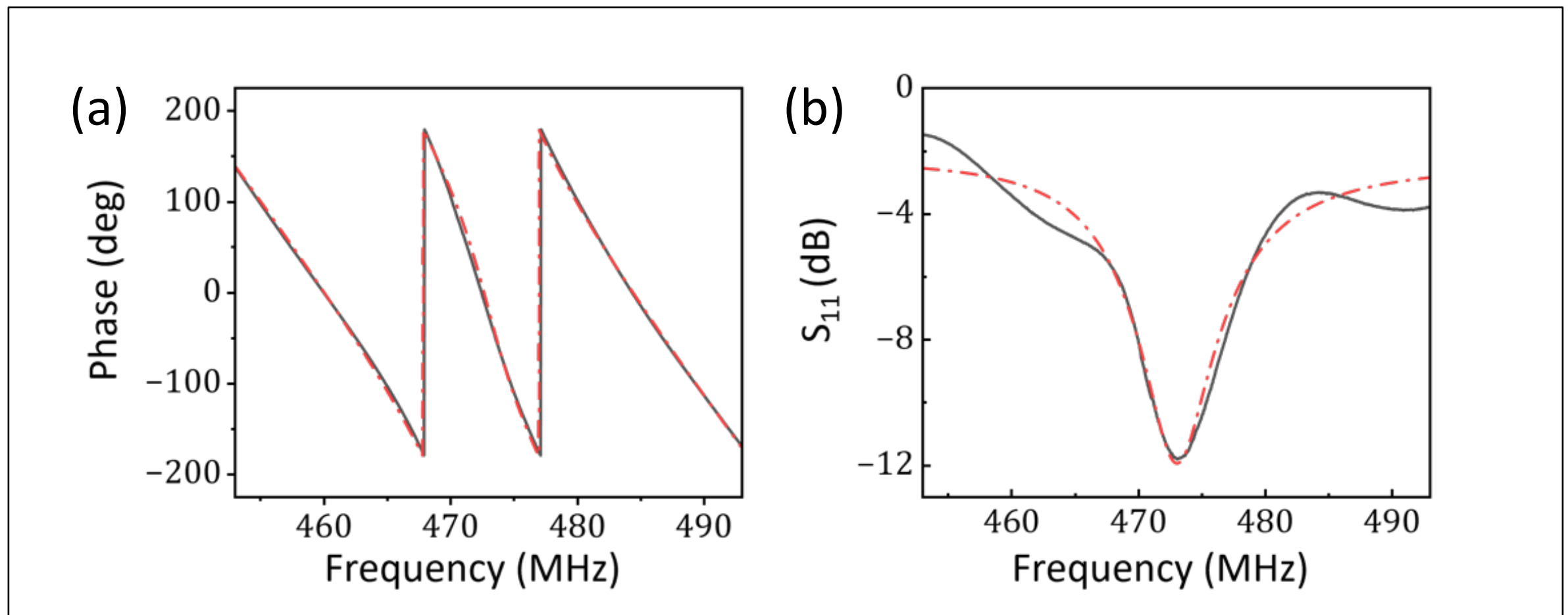

Figure S2: The S11 phase (a) and magnitude (b) of the device around its fundamental resonance at $P_{chip} = -57$ dBm. The solid black curve is the measured data and the red dotted curve corresponds to the circle fit.

The average energy stored in the resonator [3] is $\tilde{n}hf_r = P_{loss}Q_i/2\pi f_r$ where $h$ is Planck's constant, $f_r = \omega_r/2\pi$, $\tilde{n}$ is the average photon number inside the cavity, and $P_{loss}$ is the energy dissipated per resonance cycle. $P_{loss}$ is the difference between the power at the input port of the resonator $P_{chip}$ and the reflected power at resonance. Hence, $P_{loss} = P_{chip}(1 - |S_{11}|^2)$. We then calculate the average intracavity photon number as

$$\tilde{n} = \frac{P_{loss} Q_i}{2\pi h f_r^2} \tag{3}$$

Taking into account the errors from the $S_{11}$ fit, plotting $\tilde{n}$ vs. $f_r$ as in the inset to Fig. 1c of the main text and fitting to a straight line gives us a self-Kerr coefficient $K$ of $(-10.20 \pm 0.35) \times 10^{-4}$ Hz per photon.

# SI-3. Frequency Comb Generation in 980 MHz resonator

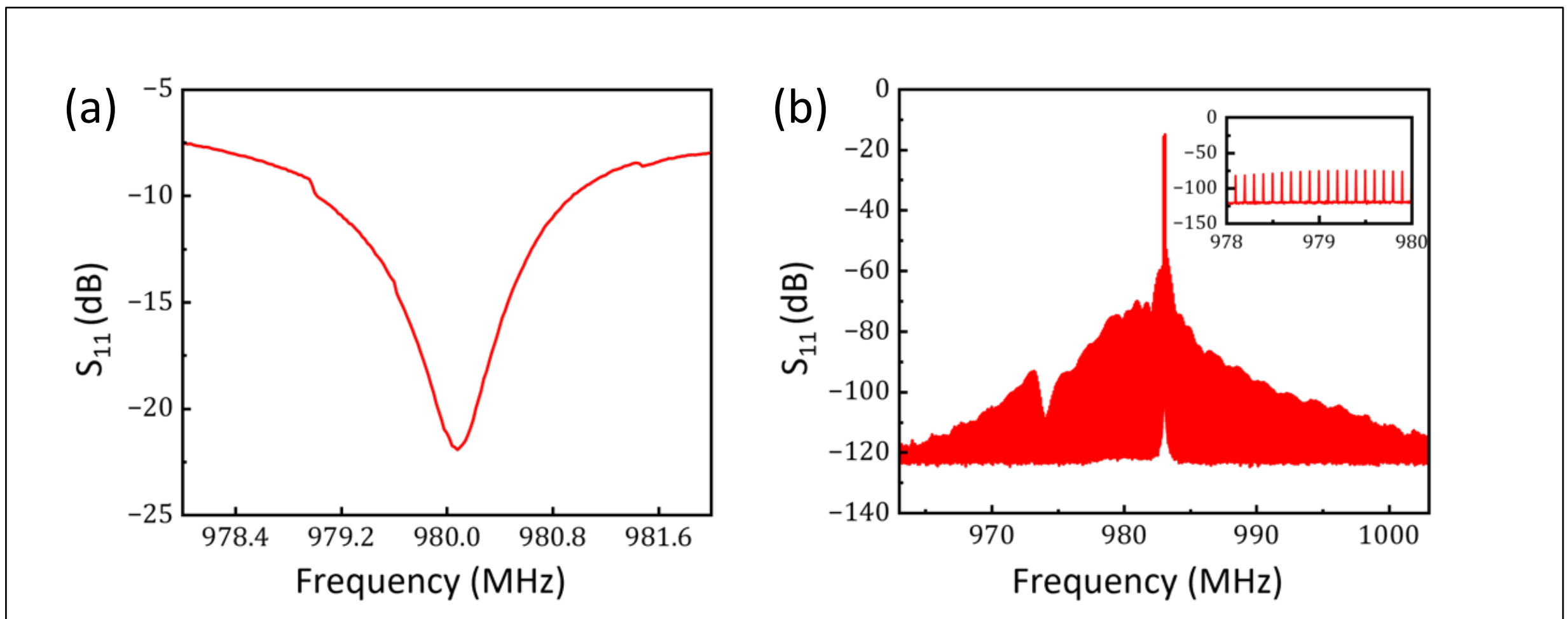

Figure S4: CPW resonator with fundamental resonance ~980 MHz (a) showing broad microwave frequency comb generation (b) when we supply two pump tones at 983 MHz and 983.01 MHz.

We have also measured microwave frequency comb generation through cascaded four-wave mixing in another CPW resonator on STO whose resonance is ~980 MHz, as shown in Fig. S4(a). By using the same measurement setup, we supply two pumps tones at 983 MHz and 983.01 MHz to get broad comb formation as shown in Fig. S4(b).

# SI-4. Phase stability of the frequency comb

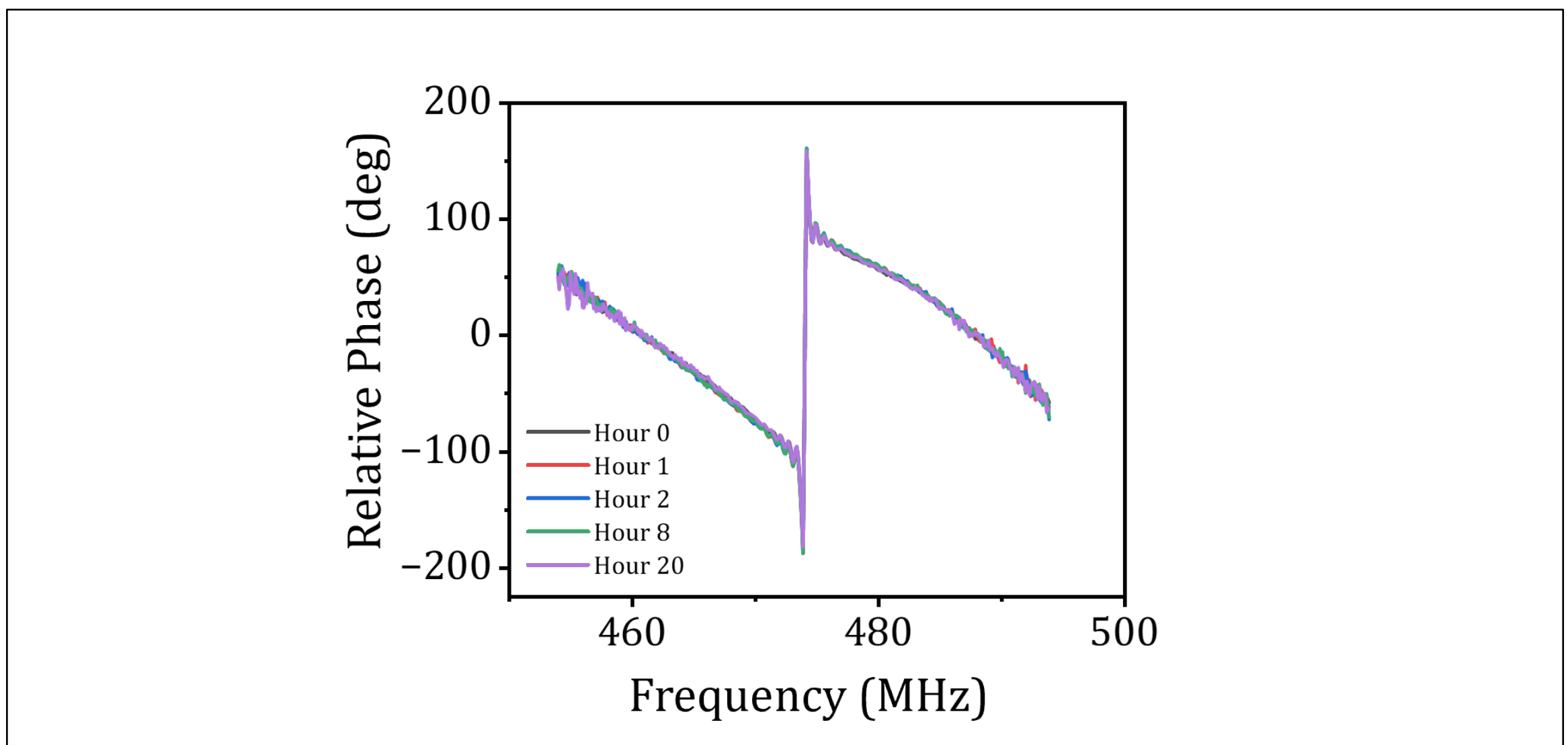

Figure S3: Long-term phase stability of the frequency comb over a period of 20 hours.

We have taken the long-term phase stability of the frequency comb over a period of 20 hours, as shown in Fig. S3. This is done by recording the time domain response of the comb in the oscilloscope (similar to Fig. 4(a) of the main text) for each hour, performing a Fourier transform of the time domain response and extracting the relative phases of the comb line. We observe that the relative phases do not vary significantly over the duration of over 20 hours as can be seen in Fig. S2.